\documentclass[aps, prd, superscriptaddress, nofootinbib, showkeys, twocolumn, showpacs, preprintnumbers, amsmath, amssymb]{revtex4-1}
\usepackage[usenames, dvipsnames]{xcolor}

\usepackage{graphicx}
\usepackage{float}

\usepackage{gensymb}
\usepackage{epstopdf}

\newcommand{\req}[1]{Eq.\,\eqref{#1}}

\begin{document}
	
	\title{Impact of Gravitational Slingshot of Dark Matter on Galactic Halo Profiles}
	
	\author{Pisin Chen}
		\email{pisinchen@phys.ntu.edu.tw}
		\affiliation{Department of Physics, National Taiwan University, Taipei 10617, Taiwan}
		\affiliation{Leung Center for Cosmology and Particle Astrophysics, National Taiwan University, Taipei 10617, Taiwan}
		\affiliation{Graduate Institute of Astrophysics, National Taiwan University, Taipei 10617, Taiwan}
		\affiliation{Kavli Institute for Particle Astrophysics and Cosmology, SLAC National Accelerator Laboratory, Stanford University, Stanford, California 94305, USA}
	
	\author{Yi-Shiou Duh}
		\email{b02202064@ntu.edu.tw}
		\affiliation{Department of Physics, National Taiwan University, Taipei 10617, Taiwan}
		\affiliation{Leung Center for Cosmology and Particle Astrophysics, National Taiwan University, Taipei 10617, Taiwan}
	\author{Lance Labun}
		\email{lance@phys.ntu.edu.tw}
		\affiliation{Department of Physics, National Taiwan University, Taipei 10617, Taiwan}
		\affiliation{Leung Center for Cosmology and Particle Astrophysics, National Taiwan University, Taipei 10617, Taiwan}
		\author{Yao-Yu Lin}
		\email{r02222073@ntu.edu.tw}
		\affiliation{Department of Physics, National Taiwan University, Taipei 10617, Taiwan}
		\affiliation{Leung Center for Cosmology and Particle Astrophysics, National Taiwan University, Taipei 10617, Taiwan}

	\date{\today}

% -----------------------------------------------------------------------
% -----------------------------------------------------------------------

\begin{abstract}
We study the impact of gravitational slingshot on the distribution of cold dark matter in early and modern era galaxies.  Multiple gravitational encounters of a lower mass dark matter particle with massive baryonic astrophysical bodies would lead to an average energy gain for the dark matter, similar to second order Fermi acceleration. We calculate the average energy gain and model the integrated effect on the dark matter profile.  We find that such slingshot effect was most effective in the early history of galaxies where first generation stars were massive, which smeared the dark matter distribution at the galactic center and flattened it from an initial cusp profile. On the other hand, slingshot is less effective after the high mass first generation stars and stellar remnants are no longer present. Our finding may help to resolve the cusp-core problem, and we discuss implications for the existing observation-simulation discrepancies and phenomena related to galaxy mergers.
\end{abstract}

\maketitle

% -----------------------------------------------------------------------
% -----------------------------------------------------------------------

%{\it Introduction} -
\section{Introduction}
In cosmological N-Body simulations, the $\Lambda$ Cold Dark Matter ($\Lambda$CDM) model performs well for the formation of the large scale structure \cite{LCDMsim}. However, on galactic scales, simulations disagree with observations on the dark matter profile and the amplitude of the central density of the dark matter halo, known respectively as the cusp-core (CC) problem and the too big to fail problem (TBTF) \cite{CC,TBTF}.  In the CC problem, simulations predict power-law-like density functions $\rho\sim r^{-1}$  for small galactocentric radius $r<R_s$ inner to a characteristic scale radius $R_s$ \cite{NFW, Moore}, while observational evidence suggests approximately constant dark matter density distribution, $\rho\sim r^0$ for $r<R_c$, with $R_c$ defining the core size \cite{De Blok}.  TBTF refers to the simulations' prediction of a central DM density significantly greater than observations allow \cite{TBTF}.
 
Many solutions have been purposed. For example, modifications of the nature of dark matter from the cold dark matter paradigm can change the distribution of dark matter on galactic scales \cite{Fuzzy, SIDM, Su, WDM}. On the other hand, within the CDM paradigm, physics missing from most simulations, such as the supernova feedback effect \cite{Mashchenko, Pontzen, Governato, Weinberg}, can help explain the formation of core structures in the halo.  However, supernova feedback does not appear to be sufficient to account for TBTF \cite{SN not solve TBTF}.

In this Letter, we discuss the effect of gravitational slingshot, which is the acceleration of a low-mass object (a dark matter particle) by gravitational scattering from a much heavier moving object (e.g. a star).  It is well-known for accelerating spacecraft, such as Voyager, to escape the Sun's gravity and travel through the outer reaches of the solar system.  With strong evidence that dark matter is primarily composed of objects with mass much smaller than typical stars \cite{Bertone, MACHO}, dark matter in galaxies must undergo slingshot, which changes their energy distribution.  We show that gravitational slingshot accelerates light DM particles and efficiently decreases the density of dark matter when and where the number density and mass of compact (baryonic) objects are high.  It applies to the class of cold dark matter models consistent with observations.

The slingshot effect leads to energy gain on average, even though single particles can be either accelerated or decelerated depending on the relative velocities and collision angles.  This average gain is simply understood considering the integrated flux: due to larger relative velocity, more head-on events occur, and the DM particle receives a larger velocity boost when encountering the star head-on.  In this respect, our proposal is similar to second-order Fermi acceleration \cite{Fermi}, which describes the motion of a charged particle gaining energy by bouncing between randomly moving interstellar magnetic clouds multiple times. 
After multiple slingshots, the accumulated energy raises the DM particle to a higher galactic orbit.  With higher stellar number density in the central region causing higher rate of slingshots, the central peak in the dark matter distribution is depleted and a cusp profile is converted over time into a core-like profile. 

The slingshot effect requires a very small mass ratio between the objects.   Consequently, the effect will be only be manifest in simulations that include (baryonic) objects much heavier than the DM simulation bodies. To date, only a few N-body simulations have implemented baryons \cite{Illustris}.

The efficiency of slingshot depends on two parameters: (1) mass of the heavy object $M_*$, since larger mass objects have larger cross-section to deflect passing DM particles; and (2) the number density of stars $n_*$, since higher stellar density increases the rate of slingshot events.  These dependencies mean the slingshot effect is more efficient in earlier stages of the universe, due to higher mass of the Population III stars around $100M_{\odot}$ \cite{Abel, Omukai, Schneider, McKee}, as well as the higher number density inferred before the supermassive black hole forms \cite{SHBH formation Review}.  With Pop III stars an obvious source for the slingshot mechanism, we will refer to the heavy compact object in the following as a star, though clearly the role of scattering centers is equally well served by other objects, including remnants of the Pop III stars.

% -----------------------------------------------------------------------
%{\it Slingshot mechanism.}
\section{Slingshot mechanism}

We consider the standard CDM paradigm, meaning the dark matter particles have nonrelativistic velocities.  Stars and other compact objects moving in a galactic gravitational potential also have nonrelativistic velocities, and hence Newtonian mechanics and gravity suffice to derive the slingshot mechanism.  A derivation in relativistic mechanics is given in \cite{Dykla04}.  

The gravitational scattering process is elastic, and slingshot occurs when the recoil of the heavier object is negligible, which is equivalent to the condition that center-of-mass frame is nearly the same as the star frame. The Galilean transformation from the galaxy frame (in which star and DM velocities $V_*,u_\chi$ are considered measured) to the center-of-mass frame has a velocity
$
V_{CM}= (m_{\chi}u_{\chi}+ m_* V_*)/(m_{\chi}+m_*) 
= V_*\left(1+\mathcal{O}(m_\chi/M_*)\right)
$
%(\frac{u_\chi}{V_*}-1)\frac{m_\chi}{M_*}+\mathcal{O}(\frac{m_\chi}{M_*})^2\right)
with corrections suppressed by powers of the mass ratio $m_\chi/M_*\lesssim 10^{-6}$, which is constrained by microlensing surveys \cite{Bertone,MACHO}.

To demonstrate the slingshot effect, consider the limiting case of head-on encounter such that the final velocity of the DM particle is opposite the initial velocity:  $\vec u_{\chi,in}\cdot \vec V_*/(|\vec u_{\chi,in}||\vec V_*|)=-1=\vec u_{\chi,out}\cdot \vec u_{\chi,in}/(|\vec u_{\chi,out}||\vec u_{\chi,in}|)$.  In the star frame the  DM particle has initial velocity $\vec u'_{in}=\vec u_{\chi,in}+\vec V_*$ and final velocity $\vec u'_{\chi,out}=\vec u_{\chi,out}+\vec V_*$.  By conservation of energy $|\vec u'_{\chi,in}|=|\vec u'_{\chi,out}|\equiv u'_\chi$.  Therefore, in the galaxy frame the velocity change is $\Delta u_{\chi}=-2u_\chi-2V_*$, so that in addition to reversing the direction of the DM particle, the encounter with the moving star boosts the velocity by $2V_*$.  The energy gain  of the DM is compensated by a small energy loss for the star, which is negligible for each individual scattering event, but can integrate to a significant effect as we discuss later.

Generalizing to any angle of incidence, the energy change each time a DM particle passes near a star is \cite{Dykla04},
\begin{equation}\label{DeltaE}
\Delta E =  m_\chi u' V_* \big(\cos\theta'_{in} -\cos\theta'_{out}\big) .
\end{equation}
This energy change is measured in the galaxy frame, and angles are defined in the coordinate system where the star velocity is $\vec V_*=-V\hat e_x$.  The incoming $\theta'_{in}$ and outgoing $\theta'_{out}=\theta'_{in}+\theta'_{def}$ angles are in the star frame for notational simplicity. As typical in the central-force problem, the azimuthal angle $\phi$ is a cyclic variable with no explicit dependence. However, the deflection angle $\theta'_{def}$ changes sign under the rotation $\phi\to\phi+\pi$. Consequently, the effect of the $\phi$ averaging is to eliminate terms odd in $\theta'_{def}$,
\begin{equation}\label{DeltaEphiavg}
\int_0^{2\pi}\!\frac{d\phi}{2\pi}\Delta E =  m_\chi (u_\chi V_* \cos\theta_{in} + V_*^2)(1-\cos\theta'_{def}).
\end{equation} 
This expression readily indicates that the average energy change is positive due to  the $V_*^2$ term, which produces the dominant scaling of the energy change.  Averaging over $\theta_{in}$, the first term yields a number $\geq 0$, as is easily proven considering $1-\cos\theta_{def}'$ is a polynomial in $\sin\theta_{in},\cos\theta_{in}$.  Stellar mass enters via the deflection angle 
\begin{equation}\label{thetadef}
%\sin(\theta'_{def}/2)=\frac{1}{1+u'^2r_p/GM},
\mathrm{csc}(\theta_{def}'/2)={1+u'^2r_p/GM},
\end{equation}
with periapse $r_p$ related to the impact parameter $b$ by
\begin{equation}
b^2=r_p^2+\frac{2GM}{u'^2}r_p\,.
\end{equation}

\section{Time evolution of DM distribution}

To see the impact of gravitational slingshot on the DM distribution, we model the halo as a sequence of concentric shells labeled by galactocentric radius $r$.   
Each shell feels the gravitational potential of the halo interior to it and the effect of gravitational slingshot.  We approximate the mean energy of particles in the shell as $-GM(r)m_\chi/2r$.  For non-virialized dark matter, this effectively underestimates the ensemble average radius as the radius of circular orbits for all particles with orbital energy $E_{orb}=-GM(r)m_\chi/2r$.  The gravitational mass $M(r)$ is contributed by both dark matter and baryons. At early times, dark matter dominates structure formation, and we set  the NFW profile as the initial condition, which also implies that dark matter, being collionless, typically is the greater fraction of mass in the central region. Therefore here we set $M(r)\simeq M_\chi(r)$, and, focusing on the inner region $r<R_s$ where $\rho (r)_{NFW} \simeq \rho_i R_s/r$, we have  $M(r)\simeq 2\pi\rho_0 R_s r^2$.  Underestimating the effective radius in the energy-radius relation for dark matter particles is thus partially offset by underestimating the gravitational potential.

We let slingshot act for a short time $dt$ and set the new energy of the shell equal to its energy at a new galactic radius $r'=r+dr$,
\begin{equation}
-\frac{Gm_\chi M_\chi(r)}{2r'}= -\frac{Gm_\chi M_\chi(r)}{2r}+ \frac{dE}{dt}dt\,.
\end{equation} 
Note that $M_\chi(r)$ remains the same at $r'$ because the amount of DM inside the shell is conserved even as the shells expand in the galaxy coordinate system.  This yields a differential equation modeling the evolution of the shell radius
\begin{equation}\label{drdt}
\frac{dr}{dt}=\frac{2r^2}{Gm_\chi M_\chi(r)}\frac{dE}{dt}.
%=\frac{ V_*^2}{\pi GR_s\rho_0}n_*(r)
\end{equation}
The differential time $dt$ should be long compared with the mean free time between slingshot events but short compared with other galactic timescales, such as gravitational relaxation time for the stars.  Provided these conditions, the rate of energy transfer to dark matter by slingshot is given by  averaging the energy change \req{DeltaE} over collision angles ($\phi,\theta$) and impact parameter
\begin{equation}\label{dEdtdefn}
\frac{dE}{dt}%\frac{\int^{b_{max}}_{b_{min}}\int^{\pi}_{0}\Delta E\,n_*\sigma v_{rel} \sin\theta_{in} d \theta_{in}2\pi bdb}{\int^{b_{max}}_{b_{min}}\int^{\pi}_{0}  \sin\theta_{in}d \theta_{in} 2\pi b db}.
=\frac{\int^{b_{max}}_{b_{min}}bdb\int d\Omega_{in} \Delta E\,n_*\sigma v_{rel} }{\int^{b_{max}}_{b_{min}}bdb\int d\Omega_{in}},
\end{equation}
Here, $d\Omega_{in}=d(\cos\theta_{in})d\phi$, $\sigma=\pi b_{max}^2$ and  $\vec v_{rel}=|\vec V_*-\vec u_\chi|=u'$ is the relative velocity. 
Without $\Delta E$ in the integrand, \req{dEdtdefn} gives the averaged rate $\Gamma=\langle n\sigma v\rangle$. The minimum impact parameter is set by the object radius; we take $b_{min}=10^{10}$\:m.  If instead the scattering center is a compact stellar remnant such as a black hole, this radius could be set smaller, but the additional range of small $b$ does not contribute significantly to the integration. We therefore ignore such possibilities, which also saves our invoking general relativistic corrections.   We set $b_{max}=\mathrm{min}(b_C,n^{-1/3})$, the smaller of either (1) the Coulomb distance $b_C$ where the stellar potential equals the mean galactic potential, or (2) the mean interstellar distance $n^{-1/3}$.

We assume the dark matter velocity distribution in the halo is isotropic, in accordance with standard halo models \cite{Kamionkowski}.  We set the DM and star velocities to an $r$-independent constant, since over the length scale of interest it varies by at most a factor of order 1.  More important is star velocity, seeing that the dominant contribution to $\Delta E$ comes from the $m_\chi V_*^2$ term in \req{DeltaEphiavg}.  For this reason, it is sensible to write
\begin{equation}\label{dEdttau}
\frac{d E}{dt}= \frac{m_\chi V_*^2}{\tau}
\end{equation}
$\tau$ defined in this way gives the mean time between $\mathcal{O}(1)$ changes in DM particle energy by scattering from stars. $\tau$ depends inversely on $n_*$, so that higher stellar density decreases the time scale for $dE/dt$.  Less manifest is its dependence on the mass $M_*$, which is embedded in the deflection angle obtained for a given impact parameter, see \req{thetadef}. 
Intuitively, larger mass extends the range of influence of the star, and large $\mathcal{O}(1)$ deflections occur more often for larger stars.  We fix $n_*(r)M_*$ as a constant to exhibit the dependence on stellar mass and we consider the total baryon density  $n_*(r)M_*$ should remain of the same order of magnitude independent of the mass of the Pop III stars.  This implies $\tau\sim M_*^{-1}$, because the scaling of the cross section $\sigma\sim M_*^2$ is compensated by the number density of stars in theproduct $n_*\sigma$ in the integrand of \req{dEdtdefn}.

% in the current galaxy with star mass around $1M_\odot$ will be around 2 orders of magnitude smaller than the early universe.}
%Intuitively, larger mass extends the range of influence of the star, and large $\mathcal{O}(1)$ deflections occur more often for larger stars.  This is borne out by the integration, and roughly $\tau\sim M_*^{-2}$, which one can anticipate by estimating the cross-section for $\mathcal{O}(1)$ deflections as $\sim\pi(GM/|u_\chi|^2)^2$.

We model the star density with the Sersic distribution $n_*(r)=n_0e^{-r/R_e}$ conventionally used in fitting surface brightness \cite{Sersic}.  In this case, the analytic solution of \req{drdt} is
\begin{equation}
\frac{r(t)}{r_i}=1+\frac{R_e}{r_i}\ln\big(1+\frac{v_0t}{R_e}e^{-r_i/R_e}\big),
\end{equation}
where $r(t)$ is the the radius at time $t$, $r_i$ the initial radius and $v_0=V_*^2/\pi G\rho_i R_s\tau$ the initial velocity of the shell at $r\to 0$.  The density after time $t$ is then
\begin{equation}
\rho(r,t)=\rho(r,0)\frac{r_i^3}{r(t)^3}= \frac{\rho_i R_s r_i^2}{r(t)^3}
\end{equation}
As we are interested in fitting the deviation from power-law behavior approaching from large $r$, we expand $r(t)$ for $r\gg R_e$ to obtain
\begin{equation}\label{rhotapprox}
\rho(r,t) \simeq \frac{ \rho_i R_s}{r+3R_e \ln\left(1+\frac{v_0t}{R_e} e^{-r/R_e}\right)},
\end{equation}
which exhibits the development of a core-like $r$-dependence that grows logarithmically with $t$.

In Table \ref{models}, we compare the timescale $\tau$, evaluated at $r=0$, in different galaxies at different epochs.  For the current Milky Way Galaxy, slingshot has negligible effect because $\tau\sim 10^{12}$ years is greater than the age of the universe.  The small rate is due to the low mass of compact (stellar) objects.
In the early universe however, the stellar mass and number density were very different. The first generation stars were composed of mainly helium and hydrogen, and state of the art star formation theory and simulation \cite{Abel,Omukai,Schneider,McKee} predicts masses a few times $100M_{\odot}$.  For a conservative estimate, we take the average stellar mass to be $100M_\odot$.

%%%%%%%%%%%%%%%%%%%%%%%%%%%%%%%%%%%%%%%%%%%%%%%%%%%%%%%%%%%%
\begin{table}\caption{Parameters of Milky Way-like (MW) and dwarf galaxy models and corresponding timescale $\tau$  from \req{dEdttau} evaluated at $r=0$.  \label{models}}
\begin{tabular}{c|c|c|c|c|c}
 & $M_*\,[M_\odot]$ & $V_*,u_\chi$\,[m/s] & $n_0$ [pc$^{-3}$] & $R_e[$kpc$]$ & $\tau$\,[years]\\ \hline
early MW & 100 & $10^5$ & $10$ & $0.2$ & $10^{10}$  \\ 
late MW & 1 & $10^5$ & $10^3$ & $0.2$ & $10^{12}$ \\ \hline
early dwarf & 100 & $2\times 10^4$ & $0.5$ & $0.1$ & $10^9$  \\ 
late dwarf & 1 & $2\times 10^4$ & $50$ & $0.1$ & $10^{11}$
\end{tabular}
\end{table}
%%%%%%%%%%%%%%%%%%%%%%%%%%%%%%%%%%%%%%%%%%%%%%%%%%%%%%%%%%%%

\begin{figure}[t]
\includegraphics[width=0.48\textwidth]{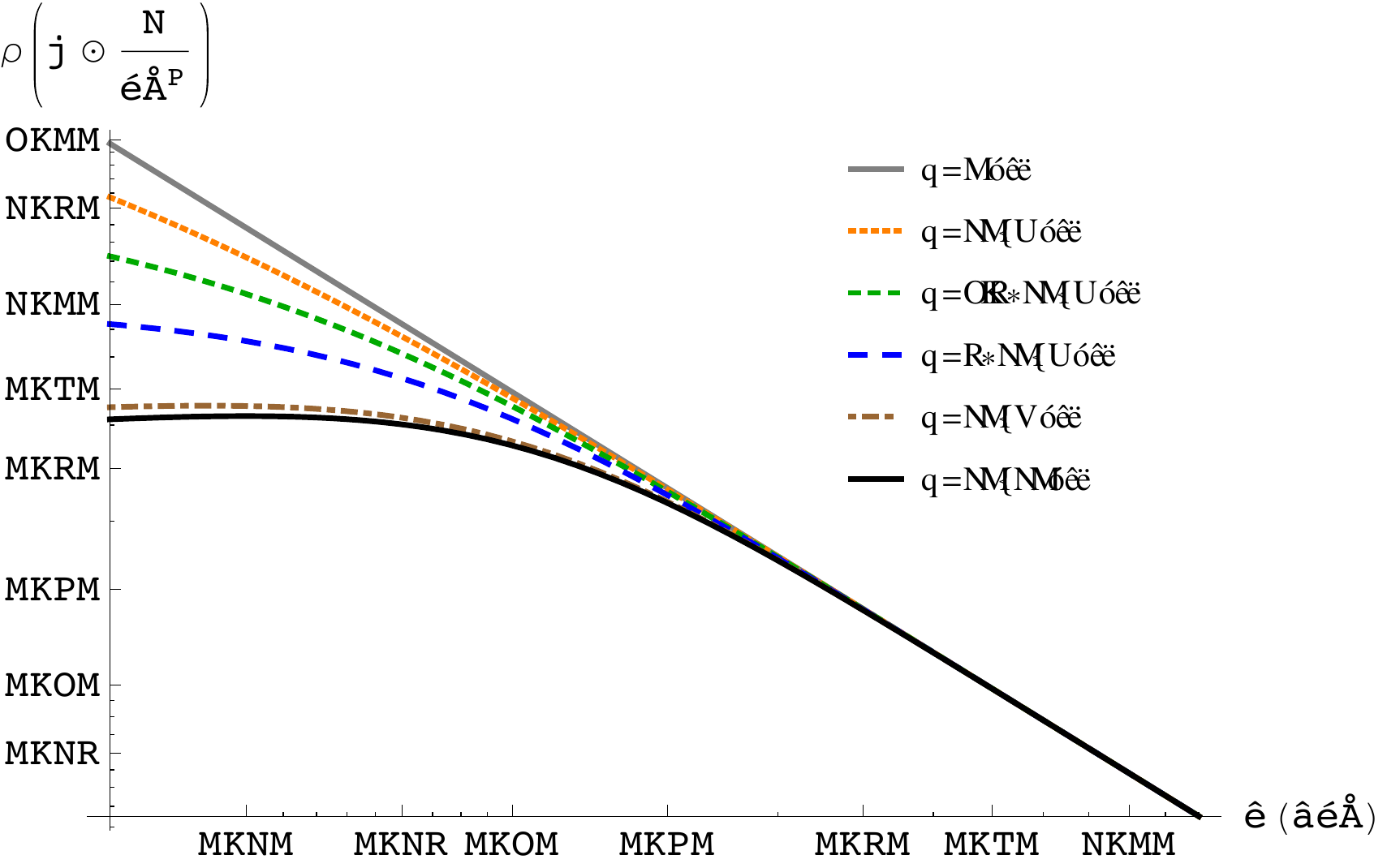}
\caption{DM density profile in the inner part of the galaxy. Solid (red) line represents the original NFW profile ($t = 0 $ yrs). Subsequent lower curves correspond to $0.1$ Gy, $0.25$ Gy, $0.5$ Gy, $1.0$ Gy, $10.0$ Gy.
\label{fig:dwarf}}
\end{figure}

For a quantitative example, we consider first parameters representing a DM-dominated dwarf galaxy in the early universe.  Here, $n_0,R_e$ are fixed by requiring the total stellar mass $\sim 10^{-2}$ the total DM halo mass and $b_{max}=10^{17}$m set by $b_C$.  We adopt the NFW parameters for the smallest halo in the catalog of \cite{NFW}, setting $\rho_i=1.5 \times 10^{-2}~ M_{\odot}/ \mathrm{pc^3} $ and  $R_s=9.2~$kpc. The density profile at different times is shown in figure \ref{fig:dwarf}.

From the evaluations of $\tau$ in Table \ref{models}, it is clear that the effectiveness of slingshot is sensitive to the star mass and number density.  The first generation stars have a short main sequence lifetime and may convert into relatively heavy remnants (e.g. a 100$M_\odot$ star becoming a $\sim 40M_\odot$ black hole \cite{Zhang} ).  With evidence of central supermassive black holes up to $z=6$ \cite{Vestergaard}, it seems likely that also the remnants have disappeared by about $10^9$ years after the Pop III stars are created. Therefore, as a rough model of the star population dynamics, we switch the distribution from an ``early universe'' parameter set to a ``late universe'' parameter set at $t=10^9$ years (=1 Gy).  The lowest line in \ref{fig:dwarf} shows that even integrating over $\sim 10^{10}$ more years, the continued action of slingshot on the dark matter profile has a small effect on the core structure established during the life cycle of Pop III stars.

\begin{figure}[t]
%\begin{center}
\includegraphics[width=0.48\textwidth]{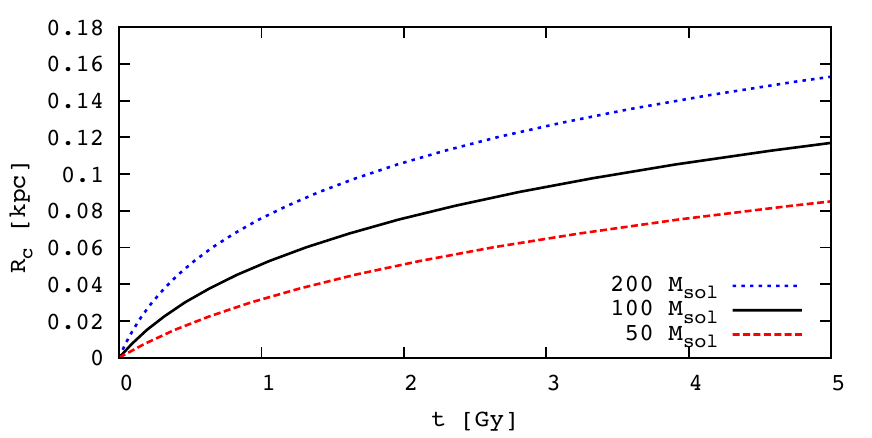}
\caption{ $R_c(t)$ in early dwarf galaxies for different fixed stellar mass. \label{fig:rc} }
%\end{center}
\end{figure}

We can immediately see that the DM density decreases most rapidly at small $r$ due to the high density of stars, and more slowly at larger $r$.  The core grows because the density decreases at larger radii more slowly, and the departure from the initial NFW becomes significant only at later time.
To show how the core growth depends on star mass, we plot in Figure \ref{fig:rc}.  We define the core radius $R_c$ as the value of $r$ such that the first term in the denominator of \req{rhotapprox} is equal to the second term.

\begin{figure}[t]
\includegraphics[width=0.48\textwidth]{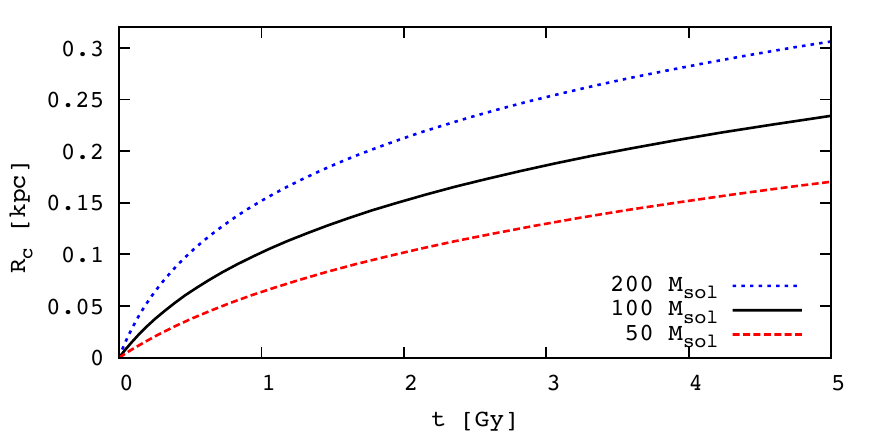}
\caption{Core radius versus time for early Milky Way-like parameters.}
\label{fig:milky}
\end{figure}

For comparison, we also model a Milky Way-like galaxy of first generation stars, see Table \ref{models} Here, we set the parameters $n_0,R_e$ such that the the total stellar mass is $10^{11} M_\odot$, according to observation \cite{SDSS}. We adopt NFW profile parameters from \cite{Nesti}, with $\rho_i=1.4 \times 10^{7}~ M_{\odot}/ \mathrm{kpc^3} $, $R_s=16.1~$kpc.  The development of the core as a function of time is shown in Figure \ref{fig:milky}.

\section{Discussion} 

By conservation of total angular momentum, the outward flow of dark matter implies loss of angular momentum by the stars.  This phenomenon is well-known in other contexts as dynamical friction, discussed already by Chandrasekhar \cite{Binney}.  Indeed, supermassive black holes may have formed and grown by dynamical friction and accretion after the first generation of stars formed \cite{SHBH formation Review}.  Our study in this way complements the work on dynamical friction, showing the implication for the dark matter distribution.  Previous results for dynamical friction are therefore consistent with our notion that the dark matter distribution was flattened due to the slingshot effect. 
Further, at late times, the back reaction of the dark matter flow on the star distribution should not be neglected and will be studied in future work. However, in general the distribution of baryons in and around stars is affected by many other processes, such as novae.  The quantitative accuracy of our results is limited rather by uncertainty in the star and dark matter distributions in early galaxies.

%The dark matter central density $\rho_i$ as well as the distribution of stars in the central kpc are not well known and estimates range over orders of magnitude introducing some theory uncertainty.

An important effect that we have not modeled here is slingshot in dark matter haloes of merging galaxies.  During mergers, slingshot occurs at even higher rate than estimated for internal galaxy dynamics because (1) the density of stars is roughly doubled and (2) the relative velocities of the merging galaxies is $10^5-10^6$ m/s, which both increases the frequency of the head-on collisions and the mean energy gain (which is maximized in head-on collisions).  Current simulations do not yet resolve the slingshot effect \cite{Boylan-Kolchin}, and its impact on the merging process will be studied in the future.

To summarize, we have shown that gravitational slingshot of light dark matter objects $m_\chi\ll M_\odot$ from heavy baryonic astrophysical bodies (stars and stellar remnants) reduces the central density of dark matter. Slingshot occurs for any cold dark matter model and is effective in the early universe with high stellar mass of Population III stars. The model we employ is designed to illuminate the important parameters and the qualitative impact. More quantitative prediction of the statistical distributions in core radius and small-$r$ profile will require at least modeling realistic galaxies with a distribution of stellar masses and lifetimes, even supernova feedback and supermassive black hole formation. Since this effect is generally missing from N-body simulations, which do not include heavier baryonic objects, we expect it helps resolve the cusp-core problem and too-big-to-fail problem between the simulations and observations.

\section*{Acknowledgments}
We would like to thank Chun-Yen Chen, Je-An Gu, Yen-Chin Ong, and Yao-Yuan Mao for helpful discussion. This work was supported by Taiwan National Science Council (TNSC) under Project No. NSC 97-2112-M-002-026-MY3 and by Taiwan National Center for Theoretical Sciences (NCTS). P.C. is in addition supported by U.S. Department of Energy under Contract No. DE-AC03-76SF00515.

%-----------------------------------------------------------------------
%----------------------------------------------------------------------

\end{document}